\documentclass[useAMS,usenatbib]{mn2e}
\usepackage{amsmath}
\usepackage{amssymb}
\usepackage{graphicx}
\def\gsim{\;\lower4pt\hbox{${\buildrel\displaystyle >\over\sim}$}\;}
\def\lsim{\;\lower4pt\hbox{${\buildrel\displaystyle <\over\sim}$}\;}
\def\grls{\;\lower4pt\hbox{${\buildrel\displaystyle >\over <}$}\;}

\title[Supermassive Black Holes in Galactic Bulges]
{Supermassive Black Holes in Galactic Bulges}
\author[Y.-Q. Lou, Y.-F. Jiang ]{Yu-Qing Lou$^{1,2,3}$
\thanks{Email: louyq@tsinghua.edu.cn; lou@oddjob.uchicago.edu}, Yan-Fei Jiang$^{1}$
\thanks{jiangyanfei1986@gmail.com; yanfeij@princeton.edu}\\
$^{1}$ Department of Physics and Tsinghua Centre for Astrophysics
(THCA), Tsinghua University, Beijing, 100084, China;
\\
$^{2}$ Department of Astronomy and Astrophysics, the University of
Chicago, 5640 South Ellis Avenue, Chicago, IL 60637, USA;\\
$^{3}$ National Astronomical Observatories, Chinese Academy of
Sciences, A20, Datun Road, Beijing, 100021, China.}
\begin{document}
\date{Accepted 2008 August 28;\ \ received 2008 August 23;\ \ in original form 2008 June 6}
 \maketitle

\begin{abstract}
Growing evidence indicate supermassive black holes (SMBHs) in a
mass range of $M_{\rm BH}$$\sim 10^6-10^{10}M_{\odot}$ lurking in
central stellar bulges of galaxies.
Extensive observations reveal fairly tight power laws of $M_{\rm
BH}$ versus the mean stellar velocity dispersion $\sigma$ of the
host stellar bulge.
Together with evidence for correlations between $M_{\rm BH}$ and
other properties of host bulges, the dynamic evolution of a bulge
and the formation of a central SMBH should be linked. In this
Letter, we reproduce the empirical $M_{\rm BH}-\sigma$ power laws
based on our recent theoretical analyses (Lou \& Wang; Wang \&
Lou; Lou, Jiang \& Jin) for a self-similar general polytropic
quasi-static dynamic evolution of bulges with self-gravity and
spherical symmetry and present a sensible criterion of forming a
central SMBH. The key result is $M_{\rm BH}={\cal L}\
\sigma^{1/(1-n)}$ where $2/3<n<1$ and ${\cal L}$ is a proportional
coefficient characteristic of different classes of host bulges. By
fitting and comparing several empirical $M_{\rm BH}-\sigma$ power
laws, we conclude that SMBHs and galactic bulges grow and evolve
in a coeval manner and most likely there exist several classes of
galactic bulge systems in quasi-static self-similar evolution and
that to mix them together can lead to an unrealistic fitting.
Based on our bulge-SMBH model, we provide explanations for
intrinsic scatter in the relation and a unified scenario for the
formation and evolution of SMBHs in different classes of host
bulges.
\end{abstract}


\section{Introduction}

It is now widely accepted that supermassive black holes (SMBHs)
within a mass range of $10^6\sim10^{10} M_{\odot}$
($M_{\odot}=2\times 10^{33}$g is the solar mass) form at the centres
of spiral and elliptical galaxies (e.g., Lynden-Bell 1969; Kormendy
\& Richstone 1995; Kormendy 2004). Observationally, SMBH masses
$M_{\rm BH}$ correlate with various properties of spiral galaxy
bulges or elliptical galaxies, including bulge luminosities (e.g.,
Kormendy \& Richstone 1995; Magorrian et al. 1998; Marconi \& Hunt
2003), stellar bulge masses $M_{\rm bulge}$ (e.g., Magorrian et al.
1998; Marconi \& Hunt 2003; H$\ddot{\rm a}$ring \& Rix 2004),
$M_{\rm BH}$ versus $M_{\rm bulge}$ relations in active (AGN) and
inactive galaxies (e.g., Wandel 1999, 2002; McLure \& Dunlop 2002),
galaxy light concentrations (e.g., Graham et al. 2001), the
S$\acute{\rm e}$rsic index (S$\acute{\rm e}$rsic 1968) of surface
brightness profile (e.g., Graham \& Driver 2007), inner core radii
(e.g., Lauer et al. 2007), spiral arm pitch angles (e.g., Seigar et
al. 2008), bulge gravitational binding energies (e.g., Aller \&
Richstone 2007) and mean stellar velocity dispersions $\sigma$
(e.g., Ferrarese \& Merritt 2000; Gebhardt et al. 2000; Tremaine et
al. 2002; Ferrarese \& Ford 2005; Hu 2008). These empirical
correlations strongly suggest a physical link between SMBHs and
their host bulges (e.g., Springel, Matteo \& Hernquist 2005; Li,
Haiman \& MacLow 2007).

Among these empirical relations, $M_{\rm BH}$ and $\sigma$
correlate tightly in a power law with an intrinsic scatter of
$\lsim 0.3$ dex (e.g., Novak et al.
2006). This
relation was explored theoretically (e.g., Silk \& Rees 1998;
Fabian 1999; Blandford 1999) before observations (e.g., Ferrarese
\& Merritt 2000; Gebhardt et al. 2000) and the emphasis was on
outflow effects for galaxies. The idea was further elaborated by
King (2003). A model of singular isothermal sphere with rotation
(Adams, Graff \& Richstone 2001) was proposed for the $M_{\rm
BH}-\sigma$ relation. This relation was also studied in a
semi-analytic model (Kauffmann \& Haehnelt 2000) with starbursts
while SMBHs being formed and fueled during major mergers.
Accretion of collisional dark matter onto SMBHs may also give the
$M_{\rm BH}-\sigma$ relation (e.g., Ostriker 2000; see Haehnelt
2004 for a review). There are also numerical simulations to model
feedbacks from SMBHs and stars on host galaxies.

Observationally, there are two empirical types of bulges:
classical bulges (spiral galaxies with classical bulges or
elliptical galaxies) and pseudobulges (e.g., Kormendy et al. 2004;
Drory \& Fisher 2007).
While SMBHs in classical bulges are formed after major mergers,
pseudobulges do not show obvious merger signatures. Interestingly,
pseudobulges also manifest a $M_{\rm BH}-\sigma$ power law yet
with a different exponent (e.g., Kormendy \& Gebhardt 2001; Hu
2008).

Self-similar dynamics of a conventional polytropic gas sphere has
been studied earlier (e.g., Yahil 1983; Suto \& Silk 1988; Lou \&
Wang 2006 (LW06) and references therein). LW06 obtained novel
self-similar quasi-static dynamic solutions which approach
singular polytropic spheres (SPS) after a long time.
Such polytropic dynamic solutions have been further generalized
and applied to study protostar formation, ``champagne flows" in H
II regions, stellar core collapse, rebound shocks and the
formation of compact stellar objects in a single fluid model
(LW06; Lou \& Wang 2007; Wang \& Lou 2007, 2008; Hu \& Lou 2008)
as well as galaxy clusters in a two-fluid model (Lou, Jiang \& Jin
2008; LJJ hereafter). Here, we take such quasi-static solutions in
a general polytropic fluid to describe long-time evolution of host
stellar bulges and the formation of central SMBHs and to establish
$M_{\rm BH}-\sigma$ power laws.

\section{A polytropic self-similar dynamic model for $M_{\rm BH}-\sigma$ power laws}

For the dynamic evolution of a stellar bulge in a galaxy, we adopt
a few simplifying assumptions. First, we treat the stellar bulge
as a spherical polytropic fluid as the typical age $\sim 10^9$ yr
of galactic bulges is long (e.g., Frogel 1998; Gnedin, Norman \&
Ostriker 1999) that they are continuously adjusted and relaxed.
Stellar velocity dispersions produce an effective pressure $P$
against the self-gravity as in the Jeans equation (e.g., Binney \&
Tremaine 1987). Secondly,
the total mass of the interstellar medium
in a galaxy is $\sim 10^7-10^8 M_{\odot}$
(e.g., Gnedin et al.
1999),
only $10^{-2}\sim 10^{-3}$ of the total bulge mass.
Although gas densities in broad and narrow line regions of AGNs are
high, the filling factor is usually small ($\sim 10^{-3}$;
Osterbrock \& Ferland 2006) and the gas breaks into clumpy clouds.
Thus gas is merged into our stellar fluid as an approximation.
Thirdly, the diameter of broad line regions of AGNs is only $\sim
0.1 $ pc (Osterbrock \& Ferland 2006) and the disc around a SMBH is
even smaller while a galactic bulge size is $\gsim 1$ kpc. We thus
ignore small-scale structures around the central SMBH of a spherical
bulge. Finally, as rotation curves of galaxies show (e.g., Binney \&
Tremaine 1987), the effect of dark matter halo in the innermost
region (around several kpcs ) of a galaxy may be neglected. So the
dark matter is not included as we study the bulge dynamics.

Hydrodynamic equations of a general polytropic bulge model with
spherical symmetry are mass conservation
\begin{eqnarray}
\frac{\partial M}{\partial t}+u\frac{\partial M}{\partial r}=0
\qquad \mbox{  and  }\qquad \frac{\partial M}{\partial r}=4\pi
r^2\rho\ ,\label{eq01}
\end{eqnarray}
or equivalently
\begin{eqnarray}
\frac{\partial\rho}{\partial
t}+\frac{1}{r^2}\frac{\partial}{\partial r}(r^2\rho u)=0\
,\label{eq02}
\end{eqnarray}
radial momentum conservation (LW06; Wang \& Lou 2008)
\begin{eqnarray}
\frac{\partial u}{\partial t}+u\frac{\partial u}{\partial
r}=-\frac{1}{\rho}\frac{\partial P}{\partial r}-\frac{GM}{r^2}\
,\label{eq03}
\end{eqnarray}
and `specific entropy' conservation along streamlines (LJJ)
\begin{eqnarray}
\left(\frac{\partial}{\partial t} +u\frac{\partial}{\partial
r}\right) \left(\frac{P}{\rho^{\gamma}}\right)=0\ ,\label{eq04}
\end{eqnarray}
where $r$ is radius and $t$ is time; $M(r,t)$ is the enclosed mass
and $u(r,t)$ is the bulk radial flow speed; $P(r,t)$ is the
effective pressure and $\rho(r,t)$ is the mass density; $\gamma$ is
the polytropic index for the stellar bulge fluid; $G$
is the gravity constant.

As the bulk flow of stellar fluid is slow, we invoke the novel
self-similar quasi-static solutions (LW06; LJJ) to model the bulge
evolution under spherical symmetry. We use a self-similar
transformation (LW06; LJJ) to solve the general polytropic fluid
equations $(\ref{eq01})-(\ref{eq04})$, namely
\begin{eqnarray}
r=K^{1/2}t^{n}x\ ,\quad \rho=\frac{\alpha(x)}{4\pi Gt^2}\
,\quad  u=K^{1/2}t^{n-1}v(x)\ ,\nonumber\\
\!\!\!\! P=\frac{Kt^{2n-4}}{4\pi G}\beta(x)\ ,\ \ \ \ \ \ \
M=\frac{K^{3/2}t^{3n-2}m(x)}{(3n-2)G}\
.\qquad\label{transformation}
\end{eqnarray}
Here, $x$ is the independent dimensionless similarity variable while
$K$ and $n$ are two scaling indices;\footnote{Here, $n$ is not the
S$\acute{\rm e}$rsic index of surface brightness profile.}
$\alpha(x)$ is the reduced mass density and $v(x)$ is the reduced
radial flow speed; $\beta(x)$ is the reduced pressure and $m(x)$ is
the reduced enclosed mass; reduced variables $\alpha,\ \beta,\ v$
and $m$ are functions of $x$ only. We require $n>2/3$ for a positive
mass.

By transformation (\ref{transformation}), we readily construct
self-similar quasi-static dynamic solutions from the general
polytropic fluid equations $(\ref{eq01})-(\ref{eq04})$ that approach
the SPS as the leading term for small $x$. Properties of such
asymptotic solutions to the leading order (LW06; Wang \& Lou 2008;
LJJ) are summarized below. Both initial ($t\rightarrow 0^{+}$) and
final ($t\rightarrow +\infty$)
mass density profiles scale as $\sim r^{-2/n}$;
accordingly, the bulge enclosed mass profile is $M\propto
r^{3-2/n}$. As $r\rightarrow 0^+$ or $t\rightarrow +\infty$, the
density and enclosed mass profiles are
\begin{eqnarray}
M=\frac{nAK^{1/n}}{(3n-2)G}r^{(3n-2)/n}\ ,\qquad\rho=\frac{A}{4\pi
G}K^{1/n}r^{-2/n}\ ,\label{masanddensity}
\end{eqnarray}
where $A\equiv \{n^{2-q}/[2(2-n)(3n-2)]\}^{1/(q+\gamma-2)}$ and
$q\equiv 2(n+\gamma-2)/(3n-2)$. For either $x\rightarrow 0^{+}$ or
$x\rightarrow +\infty$, the reduced velocity $v\rightarrow 0$,
which means at a time $t$, for either $r\rightarrow0^+$ or
$r\rightarrow+\infty$ the flow speed $u\rightarrow 0$, or at a
radius $r$, when $t$ is short or long enough, the radial flow
speed $u\rightarrow 0$. Our model describes a self-similar bulge
evolution towards a nearly static configuration after a long time
lapse, appropriate for galactic bulges at the present epoch.

As the effective pressure $P$ results from stellar bulge velocity
dispersion, we readily derive the mean velocity dispersion
$\sigma$ in a bulge. By specific entropy conservation along
streamlines, we relate $P$ with $\rho$ and $M$ (LJJ) and derive
the $P$ profile from our quasi-static solutions.
We then take the local stellar velocity dispersion as
$\sigma_L(r,t)=(\gamma P/\rho)^{1/2}$.
The asymptotic expression when $t\rightarrow +\infty$ for the local
stellar velocity dispersion in our model is therefore
\begin{eqnarray}
\sigma_L(r)=\gamma^{1/2}K^{1/(2n)}n^{q/2}A^{(q+\gamma-1)/2}r^{(n-1)/n}\
.\label{sigmaL}
\end{eqnarray}
To compare with observations, we derive the spatially averaged
stellar velocity dispersion $\sigma$ within the bulge.
The bulge boundary is taken as the radius $r_{\rm c}$ where $\rho$
drops to a value $\rho_c$ indistinguishable from the environment.
Practically, the mean velocity dispersion $\sigma$ is usually
estimated within the half-light radius which is less than the outer
bulge photometric radius. As $\sigma_L$ decreases with increasing
$r$ by equation (\ref{sigmaL}), the mean $\sigma$ would be higher
within the half-light radius. In principle, the environmental
density for different bulges is not the same and it is difficult to
give the density exactly. Here, we take a reasonable critical
density $\rho_c$ as the criterion to define the boundary of a bulge,
which means when the density of a bulge drops to the critical
density $\rho_c$, we define the domain within this radius $r_c$ as
the bulge. For a class of bulges with the same $n$, $\rho_c$ is
regarded as a constant. So the difference in $\rho_c$ reflects
different environments of bulges with different parameter $n$. One
can readily show that within $r_c$
\begin{eqnarray}
\sigma&=&\frac{3}{4\pi r_c^3}\int_0^{r_c}\sigma_L(r)4\pi
r^2dr
\nonumber\\
&=&\Big[3n^{1+q/2}\gamma^{1/2}/(4n-1)\Big](4\pi
G\rho_c)^{(1-n)/2}A^{3nq/4}K^{1/2}\nonumber\\
&\equiv& {\cal Q}K^{1/2}\ .\label{sigma}
\end{eqnarray}
\begin{figure}
\includegraphics[height=6cm]{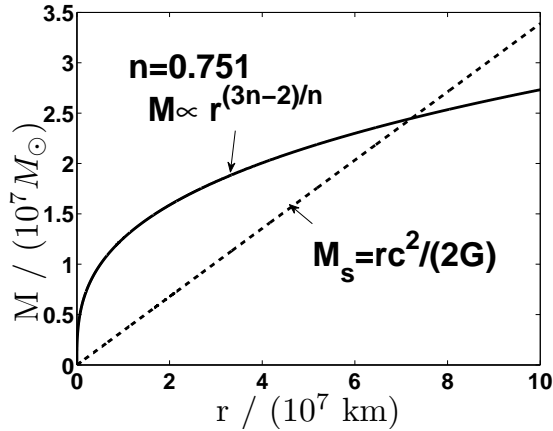}\\
\caption {The criterion of forming a central SMBH in a self-similar
quasi-static bulge evolution in a general polytropic formulation
(LW06; LJJ). The enclosed mass power law is $M\propto r^{0.337}$
with $n=0.751$ (solid curve). Meanwhile, we draw a straight dashed
line $M_{\rm s}=f rc^2/(2G)$ with $f=1$ for the mass of the central
SMBH versus the Schwarzschild radius $r$. Here at $r_{\rm
s}=7.2\times10^7$ km, the straight line intersects the enclosed mass
power-law curve for a $2.45\times 10^7M_{\odot}$ SMBH. The region
within $r_s=7.2\times 10^7$ km is the SMBH in this example. A
self-similar general polytropic quasi-static solution of $n<1$ can
thus form a SMBH at the stellar bulge centre by this
criterion.}\label{mass}
\end{figure}
This relation is particularly satisfying on the intuitive ground. A
SMBH forms at the centre of a galactic bulge that evolves in a
self-similar quasi-static manner. Such a SMBH was formed by the core
collapse of collections of stars and gas towards the bulge centre
and grows rapidly by matter accretion at an earlier phase
(e.g., Lynden-Bell 1969; Hu et al. 2006; see Haehnelt 2004 and
references therein for a review of the joint formation of SMBHs
and galaxies). As the growth timescale for SMBHs is only $\sim
10^5$ yr, our quasi-static solutions describe the relatively
quiescent phase of galactic bulges after the formation of SMBHs as
a longer history of a bulge evolution. The stellar fluid made up
of stars and condensed gas clouds has a slow bulk flow speed
towards the central SMBH, sustaining a reservoir of mass accretion
for the circumnuclear torus and/or disc on smaller scales.
Besides, there have been observational evidences to show that
stars torn up by the tidal force of the SMBH may account for the
observed X-rays at the centre of bulges (e.g., Zhao, Haehnelt \&
Rees 2002; Komossa 2004; Komossa et al. 2004; Komossa et al.
2008), which presents a scenario of how our stellar fluid being
accreted by a central SMBH.

We now introduce the criterion of forming a SMBH in a spherical
stellar bulge. A SMBH mass $M_{\rm BH}$ and
its Schwarzschild radius $r_{\rm s}$ are related by $M_{\rm BH}=f
r_{\rm s}c^2/(2G)$ where $c$
is the speed of light and $f$ is an adjustable factor of order
unity. According to equation (\ref{masanddensity}), the mass
enclosed within the radius\footnote{In the domain of bulges and
SMBHs, these asymptotic expressions represent a very good
approximation of the exact solution.} $r$ is $M\propto r^{3-2/n}$.
At a certain radius $r=r_{\rm s}=[(3n-2)f
c^2/(2nAK^{1/n})]^{n/(2n-2)}$
where it happens\footnote{We may take the last stable orbit
as the cutoff radius or a mildly relativistic case of $M_{\rm BH}=f
r_{\rm s}c^2/(2G)$ with $0.2\lsim f\lsim 3$ to form central SMBHs,
and this would not alter our results significantly.} $M=r_{\rm
s}c^2/(2G)$, a SMBH forms.
Only those quasi-static bulges with $n<1$ can form central SMBHs
as shown in Figure \ref{mass}; we derive
$M_{\rm BH}
=[f c^2/(2G)][(3n-2)f c^2/(2nA)]^{n/(2n-2)}K^{1/(2-2n)}$ and the
power law below
\begin{eqnarray}
M_{\rm BH}= \Bigg(\frac{nA}{3n-2}\Bigg)^{n/(2-2n)}
\Bigg(\frac{2}{f c^2}\Bigg)^{(3n-2)/(2-2n)}\nonumber\\
\times\frac{{\cal Q}^{1/(n-1)}}{G}\ \sigma^{\
1/(1-n)}\equiv\mathcal{L}\ \sigma^{\ 1/(1-n)}\ ,\label{relation}
\end{eqnarray}
where the coefficient $\mathcal{L}$ depends on $f c^2$, $G$, $n$,
$\gamma$, $\rho_c$, and the exponent $1/(1-n)>3$ because of the
requirement $2/3<n<1$. Other researchers have their own criteria,
which differ from ours, to define the black holes in their models to
give a $M_{\rm BH}-\sigma$ relation (e.g., Adams et al.
2001).

\begin{figure}
\includegraphics[height=6cm]{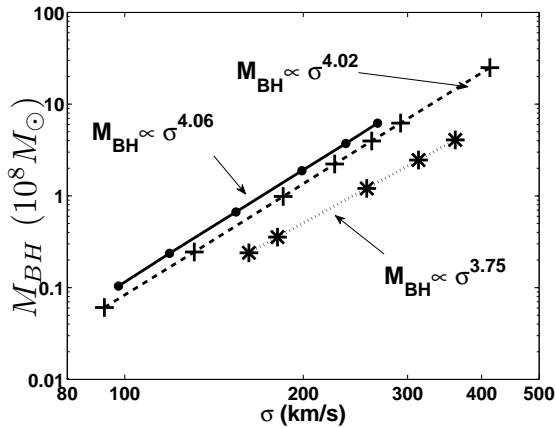}\\
\caption {Power-law $M_{\rm BH}-\sigma$ relations in our general
polytropic model for quasi-static self-similar bulge evolution and
the formation of a central SMBH. Three indices $n=0.7537$, $0.7512$,
$0.7333$ are adopted for three different classes, namely, solid line
(Hu 2008), dashed line (Tremaine et al. 2002), dotted line (Gebhardt
et al. 2000), of $M_{\rm BH}-\sigma$ power law
(\ref{relation}).
Given $n$, we calculate the SMBH mass $M_{\rm BH}$ and the mean
stellar velocity dispersion $\sigma$ for a certain $K$ value in
self-similar transformation (\ref{transformation}). A bulge has
different mean velocity dispersion $\sigma$ for different $K$
values. Each bulge-SMBH system is represented by a point in this
plot. All such systems of same $n$ lie on a straight line, while
systems of different $n$ values correspond to different lines in our
model.}\label{msigmarelation}
\end{figure}

While the $M_{\rm BH}-\sigma$ power law is very tight with
intrinsic scatter $\le 0.3$ dex for SMBHs and host bulges (e.g.,
Novak \& Faber 2006), such scatter is large enough to accommodate
different exponents in the $M_{\rm BH}-\sigma$ relation (Ferrarese
\& Merritt 2000; Gebhardt et al. 2000; Tremaine et al. 2002). By
relation (\ref{relation}), we have a natural interpretation for
intrinsic scatters in the observed $M_{BH}-\sigma$ power law. In
our model, all bulges with the same $n$ lie on a straight line
with the exponent $1/(1-n)$ as shown in Figure
\ref{msigmarelation}. For a fixed $n$, different bulges are
represented by different $K$ values in transformation
(\ref{transformation}), leading to different $M_{\rm BH}$ and
$\sigma$. However, for bulges with different $n$ values, they lie
on different lines. For elliptical galaxies or bulges in spiral
galaxies, they appear to eventually take the self-similar
evolution described above with a certain $n$ value. But
pseudobulges may take on different $n$ values. Observationally, we
cannot determine a priori the specific $n$ value for a bulge but
simply attempt to fit all bulges with a single exponent, which
then leads in part to intrinsic scatter in the observed $M_{\rm
BH}-\sigma$ power law.

To show this, we fit three published $M_{\rm BH}-\sigma$ power laws
in Figure \ref{msigmarelation}. The first one is $M_{\rm
BH}=1.2\times10^8M_{\odot}(\sigma/200\hbox{ km s}^{-1})^{3.75}$
given in Gebhardt et al. (2000) with our parameters $\{n,\ \gamma,\
\rho_c\}$ being $\{0.733,\ 1.327,\ 0.47\ M_{\odot}
\hbox{pc}^{-3}\}$; and the five points (asterisks $*$) correspond to
$K=\{0.8,\ 1,\ 2,\ 3,\ 4 \}\times10^{23}$ cgs unit and $r_c=0.73,\
0.82,\ 1.15,\ 1.41,\ 1.63$ kpc. The second one is $\log (M_{\rm
BH}/M_{\odot})=8.13+4.02\log (\sigma/ 200\hbox{ km s}^{-1})$ given
in Tremaine et al. (2002) with our parameters $\{n,\ \gamma,\
\rho_c\}$ being $\{0.7512$, 1.330, $0.0122\ M_{\odot}
\hbox{pc}^{-3}\}$; and the seven points (plus signs $+$) correspond
to $K=\{1,\ 2,\ 4,\ 6,\ 8,\ 10,\ 20 \}\times 10^{22}$ cgs unit and
$r_c=2.92,\ 4.13,\ 5.83,\ 7.14,\ 8.25,\ 9.22,\ 13.04$ kpc. The third
one is $\log (M_{\rm BH}/M_{\odot})=8.28+4.06\log (\sigma/ 200\hbox{
km s}^{-1})$ given in Hu (2008) with our parameters $\{n,\ \gamma,\
\rho_c\}$ being $\{0.7537,\ 1.332$, $0.00364\ M_{\odot}
\hbox{pc}^{-3}\}$; and the six points (solid dots) correspond to
$K=\{0.6,\ 0.9,\ 1.5,\ 2.5,\ 3.5,\ 4.5 \}\times10^{23}$ cgs unit and
$r_c=5.74,\ 7.03,\ 9.08,\ 11.72,\ 13.87,\ 15.73$ kpc. Clearly, to
fit all these points in Figure \ref{msigmarelation} with a single
power law, we would get a different result with higher intrinsic
scatter. In fact, there is yet another $M_{\rm BH}-\sigma$ relation
given in Ferrarese \& Ford (2005), namely $\log (M_{\rm
BH}/M_{\odot})=8.22+4.86\log (\sigma/200\hbox{ km s}^{-1})$ (also
Ferrarese \& Merritt 2000). If we were to fit this relation
according to our model (e.g., $n=0.7942$,
$\rho_c=1.46\times10^{-7}M_{\odot} \hbox{pc}^{-3}$, $r_c=3.25$ Mpc),
the critical mass density $\rho_c$ would be too small and the bulge
size for a certain average velocity dispersion would be too large.
So it seems that our model favours multiple power laws contained in
the available data. In the three more sensible fitting examples
above, bulge inflow speeds of stellar fluid are slow ($\sim 0.1-1$
km s$^{-1}$), an evolution feature of our self-similar quasi-static
solutions. Near the SMBH boundary $r_{\rm s}$, the inflow rest
mass-energy flux falls in the range of $10^{40}-10^{45}\hbox{ erg
s}^{-1}$ in these examples, sufficient to supply the observed X-ray
luminosities (Komossa et al. 2008). There can be outgoing accretion
shocks around a SMBH in these inflows. As the age of galactic bulges
is so long ($\sim10^{9}$ yr) (Frogel 1998) that such shocks should
have already gone outside bulges and dispersed or merged into
surroundings.

Besides the $M_{\rm BH}-\sigma$ relation,
observations reveal $M_{\rm BH}\propto M_{\rm bulge}^{1.12}$ with
$M_{\rm bulge}$ being the stellar bulge mass (e.g., H$\ddot{\rm
a}$ring \& Rix 2004 and references therein) and $M_{\rm BH}\propto
E_{\rm g}^{0.6}$ with $E_{\rm g}$ being the absolute value of the
bulge gravitational binding energy (e.g., Aller \& Richstone 2007).
Using our criterion of forming a SMBH and the bulge radius $r_c$, we
derive a power law between $M_{\rm BH}$ and $M_{\rm bulge}$ as
$M_{\rm BH}\propto M_{\rm bulge}^{1/(3-3n)}$ according to equation
(\ref{masanddensity}). For $n=0.75$, our result leads to relations
in Adams et al.
(2001) but for a nonisothermal general SPS. The bulge gravitational
binding energy, without contributions from dark matter halo and a
disc as in Aller \& Richstone (2007), is $E_{\rm
g}\approx\int_0^{r_c}{GM}\rho 4\pi rdr$. For self-similar
quasi-static dynamic solutions in general polytropic fluid model, we
obtain $M_{\rm BH}\propto E_{\rm g}^{1/(5-5n)}$.

As another class of bulges, pseudobulges are thought to have formed
without merging in contrast to classical bulges. Pseudobulges again
follow a $M_{\rm BH}-\sigma$ power law (Hu 2008), i.e.,
$\log (M_{\rm BH}/M_{\odot})=7.5+4.5\log (\sigma/ 200\hbox{ km
s}^{-1})$. Pseudobulges may take a different self-similar
quasi-static evolution for a different $n$. For their different
formation history, they show a different $M_{\rm BH}-\sigma$ power
law as observed. For $\{n,\ \gamma,\ \rho_c\}$ being $\{0.7778,\
1.34,\ 0.000426\ M_{\odot}\hbox{pc}^{-3}\}$, we can fit the
empirical power law for pseudobulges; e.g., with $K=6\times 10^{22}$
cgs unit, we have $\sigma=124\hbox{ km s}^{-1}$ and $r_c=5.742$ kpc
for elliptical galaxies.

\section{Conclusions and Discussion}

On the basis of a self-similar quasi-static dynamic evolution of a
general polytropic sphere (LW06, LJJ), we establish $M_{\rm
BH}=\mathcal{L}\ \sigma^{\ 1/(1-n)}$ with $2/3<n<1$ by equation
(\ref{relation}). Our first conclusion is $1/(1-n)>3$ which appears
consistent with observations so far. Secondly, uncertainties in the
formation criterion of a SMBH [i.e., factor $f$ in $M_{\rm BH}=f
r_sc^2/(2G)$] and in the choice of $r_c$ and thus $\rho_c$ will not
change the form of equation (\ref{relation}) and the $n$ value but
will only affect the value of ${\cal L}$. Thirdly, the tight $M_{\rm
BH}-\sigma$ power laws and other relations among the SMBH mass
$M_{\rm BH}$ and known properties of host stellar bulges strongly
suggest coeval growths of SMBHs and galactic bulges (e.g., Page et
al. 2001; Haehnelt 2004; Kauffmann \& Haehnelt 2000; Hu et al.
2006).
Fourthly in our model, while forming a SMBH at the bulge centre
(e.g., by core collapse of gas and stars or by merging), the
spherical general polytropic bulge evolves in a self-similar
quasi-static phase for a long time.
We then reproduce well-established empirical $M_{\rm BH}-\sigma$
power laws. Different energetic processes appear to give rise to
different scaling index $n$ values, which finally determines the
slope of the $M_{\rm BH}-\sigma$ relations in a logarithmic
presentation.

Besides classical bulges and pseudobulges, there are also `core'
elliptical galaxies (i.e., those with apparent `cores' of relatively
flat brightness; Lauer et al. 1995; Hu 2008), thought to have formed
by `dry' mergers (i.e., almost without gas). A steeper $M_{\rm
BH}-\sigma$ relation exists in these galaxies as compared to that
for classical bulges (e.g., Lauer et al. 1995; Laine et al. 2003;
Lauer et al. 2007).

This can also be accommodated in our unified scenario that all hosts
of SMBHs may finally evolve into self-similar quasi-static phase
with different scaling parameters (i.e., different index $n$ for the
slope and different $\rho_c$ for the normalization of the $M_{\rm
BH}-\sigma$ relation). While these bulges can be of quite different
kinds in galaxies, they have these similar tight relations and we
thus provide a unified self-similar dynamic framework to model the
relatively quiescent evolution phase of SMBH host bulges and the
growth of SMBH masses. As the observed $M_{\rm BH}-\sigma$ relation
for classical bulges is tight, the elliptical galaxies and spiral
galaxies appear to take on close $n$ values for merging processes.

In our model, $n$ is a key scaling index to determine the exponent
of the $M_{\rm BH}-\sigma$ power law. The smaller the value of $n$
is, the steeper the density profile is and the smaller the index
of the $M_{\rm BH}-\sigma$ relation is. If SMBHs are formed by
collapse of stars and gas and a less steeper density distribution
may provide a more effective mechanism to form SMBHs, then we
conclude that for a certain value of velocity dispersions,
\footnote{The mean velocity dispersion $\sigma={\cal Q}K^{1/2}$ is
not sensitive to $n$.}
the smaller the mass of an initially formed SMBH is, the smaller
the value of $n$ is.

After a long lapse, our quasi-static solutions approach the static
SPS solution as the leading term that is independent of the
timescale. So all the described relations here are nearly
independent of time, as long as the systems have evolved for a long
enough time. It is not obvious to decide when the host bulges began
to take the described self-similar evolution. But even if we take
different times in our model, our results would remain largely the
same. The only difference is that for an early time, we may have a
chance to observe accretion shocks within the region of stellar
bulges. Such shocks are characterized by a rapid inner density rise
of several times and a rapid inner rise of stellar velocity
dispersions depending on the strength of accretion shocks.



\section*{ACKNOWLEDGEMENTS}

We thank referee A. Wandel for suggestions to improve the
manuscript. This work was supported in part by Tsinghua Centre for
Astrophysics (THCA),
by the National Natural Science Foundation of China (NSFC) grants
10373009 and 10533020, by the National Basic Science Talent
Training Foundation
(NSFC J0630317) and by the National Scholarship from the Ministry
of Education
at Tsinghua University, and by the Yangtze Endowment and the SRFDP
20050003088 from the Ministry of Education
at Tsinghua University.
The kind hospitality of Institut f\"ur Theoretische Physik
und Astrophysik der Christian-Albrechts-Universit\"at Kiel
is gratefully acknowledged.


\begin{thebibliography}{120}

\bibitem{adams2001}
Adams F. C., et al.,
2001, ApJ, 551, L31

\bibitem{aller2007}
Aller M. C., Richstone D. O., 2007, ApJ, 665, 120

\bibitem{binneytremaine1987}
Binney J., Tremaine S., Galactic Dynamics (Princeton University
Press, New Jersey, 1987)

\bibitem{blandford1999}
Blandford R. D., 1999, in ASP Conf. Ser. 182, Galaxy Dynamics,
eds. D. R. Merritt, M. Valluri, J. A. Sellwood (San Francisco:
ASP), 87

\bibitem{droryfisher2007}
Drory N., Fisher D. B., 2007, ApJ, 664, 640

\bibitem{fabian1999}
Fabian A. C., 1999, MNRAS, 308, L39

\bibitem{ferrarese2000}
Ferrarese L., Merritt D., 2000, ApJ, 539, L9

\bibitem{ferrarese2005}
Ferrarese L., Ford H., 2005, SSR, 116, 523

\bibitem{frogel1988}
Frogel J. A., 1988, ARA\&A, 26, 51

\bibitem{gebhardtetal2000}
Gebhardt K. et al., 2000, ApJ, 539, L13

\bibitem{gnedin1999}
Gnedin N. Y., et al.,
1999, AIPC, 470, 48

\bibitem{graham2001}
Graham A. W., et al.,
2001, ApJ, 563, L11

\bibitem{graham2007}
Graham A. W., Driver S. P., 2007, ApJ, 655, 77

\bibitem{haehnelt2004}
Haehnelt M. G., Coevolution of Black Holes and Galaxies, from the
Carnegie Observatories Centennial Symposia, 405 (Cambridge
University Press, Cambridge 2004)

\bibitem{haringrix2004}
H$\ddot{{\rm a}}$ring N., Rix H.-W., 2004, ApJ, 604, L89

\bibitem{hu2008}
Hu J., 2008, MNRAS, 386, 2242

\bibitem{hushenlouzhang2006}
Hu J., et al.,
2006, MNRAS, 365, 345

\bibitem{hulou2008}
Hu R. Y., Lou Y.-Q., 2008 MNRAS (arXiv0808.2090H)

\bibitem{kauffmann2000}
Kauffmann G., Haehnelt M., 2000, MNRAS, 311, 576

\bibitem{king2003}
King A., 2003, ApJ, 596, L27

\bibitem{komossa2004}
Komossa S., 2004, {\it Proceedings IAU Symposium}, 222, 45

\bibitem{komossaetal2004}
Komossa S., et al.,
2004, ApJ, 603, L17

\bibitem{komossa2008}
Komossa S., et al.,
2008, ApJ, 678, L13

\bibitem{kormendy1995}
Kormendy J., Richstone D., 1995, ARA\&A, 33, 581

\bibitem{kormendy2004}
Kormendy J., 2004, in Carnegie Obs. Astrophys. Ser. Vol. 1,
Coevolution of Black Holes and Galaxies, ed. L. C. Ho (Cambridge:
Cambridge University Press), 1

\bibitem{kormendy2001}
Kormendy J., Gebhardt K., The 20th Texas Symposium on Relativistic
Astrophysics, 363 (2001)

\bibitem{laineetal2003}
Laine S., et al.,
2003, AJ, 125, 478

\bibitem{laueretal1995}
Lauer T. R., et al.,
1995, AJ, 110, 2622

\bibitem{laueretal2007}
Lauer T. R. et al., 2007, ApJ, 662, 808

\bibitem{lietal2007}
Li Y., Haiman Z., MacLow M. M., 2007, ApJ, 663, 61

\bibitem{loujiangjin2008}
Lou Y.-Q., Jiang Y. F., Jin  C. C., 2008, MNRAS, 386, 835

\bibitem{louwang2006}
Lou Y.-Q., Wang W. G., 2006, MNRAS, 372, 885 (LW06)

\bibitem{louwang2007}
Lou Y.-Q., Wang W. G., 2007, MNRAS, 378, L54

\bibitem{lyndenbell1969}
Lynden-Bell D., 1969, Nature, 223, 690

\bibitem{magorrianetal1998}
Magorrian J. et al., 1998, AJ, 115, 2285

\bibitem{marconihunt2003}
Marconi A., Hunt L. K., 2003, ApJ, 589, L21

\bibitem{McLureDunlop2002}
McLure R. J., Dunlop J. S., 2002, MNRAS, 331, 795

\bibitem{novak2006}
Novak G. S., Faber S. M., Dekel  A., 2006, ApJ, 637, 96

\bibitem{osterbrock2006}
Osterbrock D. E., Ferland G. J., Astrophysics of Gaseous Nebulae
and Active Galactic Nuclei (University Science Books, 2nd Edition,
2006)

\bibitem{ostriker2000}
Ostriker J. P., 2000, PRL, 84, 5258


\bibitem{pageetal2001}
Page M. J., et al.,
2001, Science, 294, 2516

\bibitem{seigar2008}
Seigar M. S., et al.,
2008, ApJ, 678, 93

\bibitem{sersic1968}
S\'ersic J.-L., 1968, Atlas de Galaxias Australes (Observatorio
Astronomico, Cordoba)

\bibitem{silkrees1998}
Silk J., Rees M. J., 1998,  A\&A, 331, L1

\bibitem{springel2005}
Springel V., et al.,
2005, MNRAS, 361, 776

\bibitem{SutoSilk1988}Suto Y.,
Silk J., 1988, ApJ, 326, 527

\bibitem{tremaine2002}
Tremaine S. et al., 2002, ApJ, 574, 740

\bibitem{Wandel1999}
Wandel A., 1999, ApJ, 519, L39

\bibitem{Wandel2002}
Wandel A., 2002, ApJ, 565, 762

\bibitem{wanglou2007}
Wang W. G., Lou Y.-Q., 2007, ApSS,
311, 363

\bibitem{wanglou2008}
Wang W. G., Lou Y.-Q., 2008, ApSS,
315, 135

\bibitem{Yahil1983}Yahil A.,
1983, ApJ, 265, 1047

\bibitem{zhao2002}
Zhao H. S., Haehnelt M. G., Rees M. J., 2002, NA, 7, 385

\end{thebibliography}
\end{document}